# *Temporal Brewster angle*

*Victor Pacheco-Peña[1] and Nader Engheta[2]*

[1]*School of Mathematics, Statistics and Physics, Newcastle University, Newcastle Upon Tyne, NE1 7RU, United Kingdom*
[2]*Department of Electrical and Systems Engineering, University of Pennsylvania, Philadelphia, PA 19104, USA*
*email: Victor.Pacheco-Pena@newcastle.ac.uk, engheta@ee.upenn.edu*

**Controlling amplitude, phase and polarization of electromagnetic waves is key for a full manipulation of wave-matter interactions. The Brewster angle is one of the important features in this context. Here, we exploit metamaterial concepts with a time-modulated permittivity to propose the temporal equivalent of the spatial Brewster angle, a concept we call *temporal Brewster angle*. We consider temporal boundaries (as the temporal equivalent of the spatial boundaries between two media) by rapidly changing the permittivity of the medium, where a wave travels, from isotropic to an anisotropic permittivity tensor. It is theoretically shown that when the incidence angle coincides with that of the *temporal Brewster angle* a *forward* (temporal transmission) wave is produced while the *backward* (temporal reflection) is eliminated. We provide a closed-form analytical expression of the *temporal Brewster angle* and demonstrate its performance both theoretically and numerically. Our findings may provide a fresh view on how to control electromagnetic wave propagation and wave-matter interactions in real time using temporal metamaterials.**

Manipulating electromagnetic (EM) waves has been an active research topic for many years in order to fully control their propagation in terms of amplitude, phase, polarization, etc. One of the simplest yet important discoveries in this context was the study of polarization of light carried out by Brewster in the early 19th century [1]. It was shown that obliquely incident unpolarized light can be linearly polarized when encountering an interface between two media of different refractive index. The angle at which this phenomenon occurs is known as Brewster's angle and can be described by the Fresnel equations for transmission and reflection of waves at the interface between two media. In this realm, *p*-polarized light impinging into non-magnetic medium 2 from non-magnetic medium 1 with an angle equals to the Brewster angle can be transmitted (and refracted) into medium 2 without suffering reflections at the interface between the two materials [2]. This phenomenon has had many implications in fields such as optics as it provided the scientific community with a full physical understanding of how light can be polarized and how such polarized light can be manipulated.

Controlling wave-matter interaction with metamaterials (and metasurfaces as the 2D version) have also become a prominent research field within the last few years as they provide an at-will tailoring of the EM response (permittivity, ε, permeability, μ) of media enabling extreme parameter values such as near-zero and negative refractive index [3,4]. Metamaterials have been proposed for new and improved applications [5-17] such as lenses and antennas [5–8], sensors [9,10], tunable metamaterials [11,12], quantum technologies [13,14], deep learning [15,16], computing [17] and even more sophisticated approaches based on the Brewster angle [18], to name a few. This demonstrates numerous possibilities and opportunities that metamaterials offer in multiple spectral ranges and various scenarios [19–23]. While most metamaterials topics have been dealt in the time-harmonic scenario (frequency domain), recent years have witnessed growing interest in spatiotemporal and temporal metamaterials for controlling EM wave propagation in four dimensions (space and time) [24]. The temporal modulation of the EM properties of media was introduced in the last century [25,26] considering a monochromatic wave traveling in a medium whose relative permittivity was rapidly changed in time from $\varepsilon_{r1}$ to $\varepsilon_{r2}$ (all values positive and larger than 1) at a time t = $t_1$. In so doing, it was shown how such temporal boundary gives rise to a forward (FW) and a backward (BW) wave (temporal equivalent of transmitted and reflected waves at the spatial interface between two media, respectively). This spatiotemporal modulation of the EM properties of media have recently exploited in exotic applications such as Fresnel drag in spatiotemporal metamaterials [27], nonreciprocity [28,29], effective medium theory [30], antireflection temporal coatings [31], frequency conversion [32], inverse prism [33], holography [34] and temporal beam steering (temporal aiming) using temporal anisotropic metamaterials [35].

Inspired by the exciting opportunities offered by spatiotemporal metamaterials and the importance of transmission of electromagnetic waves without reflections, in this Letter we propose a concept that we call *temporal Brewster angle* as the temporal analogue of the conventional spatial Brewster angle. In our recent work [35], we have demonstrated, theoretically and numerically, how the direction of energy propagation of a *p*-polarized EM wave can be modified in time (while the direction of the phase flow stays unchanged) when inducing a temporal boundary via a rapid temporal change of the permittivity of the medium ε(t) from an isotropic $\varepsilon_{r1}$ to an anisotropic permittivity tensor $\overline{\overline{\varepsilon_{r2}}} = \{\varepsilon_{r2x}, \varepsilon_{r2z}\}$ (all relative values are larger than unity, assuming no dispersion since we consider that any material resonance frequencies are much larger than the frequency of operation). In the present work, we introduce a novel concept in such isotropic-to-anisotropic temporal variation of permittivity in which the FW wave occurs without any BW wave for a special oblique incidence angle for the *p*-polarized EM wave. It is shown how such incidence angle ($\theta_1$), which we call *temporal Brewster angle* ($\theta_{tB}$), is determined and derived analytically in terms of the parameters involved.

Consider the conventional spatial scenario depicted in Fig. 1(a). A monochromatic *p*-polarized EM wave travels in medium 1 ($\varepsilon_{r1}$) towards medium 2 ($\varepsilon_{r2}$) with an incidence angle $\theta_i$ (we consider non-magnetic materials having $\mu_r$ = 1). Obviously since there is an impedance mismatch between the two media, a transmitted and reflected waves are produced, traveling in medium 2 and 1 with angles $\theta_t$ and $\theta_r$, respectively. It is well known that if the incidence angle is selected such that $\theta_t + \theta_r$ = 90° no reflected wave is produced for this *p*-polarization as it corresponds to the well-known Brewster angle defined as $tan(\theta_i) = \sqrt{\varepsilon_{r2}}/\sqrt{\varepsilon_{r1}}$ [2] (see schematic representation in Fig. 1(b)).

Let us now consider the temporal boundary shown in Fig. 1(c) where a monochromatic *p*-polarized EM wave is traveling in a spatially unbounded medium (for times t < $t_1$). If a temporal boundary is induced at t = $t_1$ via a rapid change of ε from isotropic $\varepsilon_{r1}$ to another isotropic value, $\varepsilon_{r2}$, for the set of FW and BW waves created at such temporal boundary: i) the wave vector ***k*** is preserved; ii) their frequency is changed from $f_1$ to $f_2$ such that $f_2 = (\sqrt{\varepsilon_{r1}}/\sqrt{\varepsilon_{r2}})f_1$ [25,26] and, interestingly iii) they travel with the same angle as before the temporal change $\varepsilon(t)$ was induced (i.e., $\theta_1 = \theta_2 = \theta_{1k,S} = \theta_{2k,S}$ with subscript $S$ denoting the direction of energy or Poynting vector $\boldsymbol{S}$). These results are expected since a sense of oblique incidence does not exist when considering a temporal boundary, where ε is modified in time from isotropic-to-isotropic, in a spatially unbounded medium. However, one may ask: is there a way to eliminate the BW wave for such *p*-polarized EM wave using temporal metamaterials? To achieve this, different techniques can be applied such as i) inducing a temporal boundary by rapidly changing both ε and μ from $(\varepsilon_{r1}, \mu_{r1})$ to $(\varepsilon_{r2}, \mu_{r2})$ such that the impedance is preserved before and after the temporal change [32] or also ii) applying an antireflection temporal coating (as the temporal analogue of the conventional antireflection coating such as quarter-wave impedance transformer) as we have recently proposed [31]. The elimination of the BW wave has also been recently proposed by using transmission lines with time-dependent parameters [36]. However, what would happen if we rapidly change in time the permittivity from isotropic ($\varepsilon_{r1}$) to anisotropic ($\overline{\overline{\varepsilon_{r2}}} = \{\varepsilon_{r2x}, \varepsilon_{r2z}\}$)? The schematic representation of such scenario is shown in Fig. 1(d). As we have shown in [35], such temporal boundary also preserves the wave vector $\boldsymbol{k}$ but it affects the direction of the energy flow ($\boldsymbol{S}$) for both FW and BW waves, which is different from the direction of the phase flow (i.e., $\boldsymbol{k}$) of the monochromatic wave before the temporal change of permittivity i.e., ($\theta_1 = \theta_{1k,S} = \theta_{2k}) \neq (\theta_{2=} \theta_{2S}$). Such change of the direction of energy $\theta_{2S}$ is a direct consequence of the change of amplitude of each component of the electric field ($E_x$ and $E_z$) due to the new anisotropic permittivity tensor $\overline{\overline{\varepsilon_{r2}}} = \{\varepsilon_{r2x}, \varepsilon_{r2z}\}$, analytically derived by us as $\theta_{2S} = tan^{-1}\left[tan(\theta_1)\left(\frac{\varepsilon_{r2z}}{\varepsilon_{r2x}}\right)\right]$ in [35]. Finally, the normalized amplitude of the FW ($\frac{E_2^+}{E_1}$) and BW ($\frac{E_2^-}{E_1}$) waves after such isotropic-to-anisotropic temporal modulation of ε can be expressed as follows [35]:

$$\frac{E_2^\pm}{E_1} = \frac{1}{2}\left[\frac{\mu_{r2y}\omega_2 \pm \mu_1\omega_1}{\mu_{r2y}\omega_2}\right]\frac{\varepsilon_{r1}\sqrt{\varepsilon_{r2z}^2 k_z^2 + \varepsilon_{r2x}^2 k_x^2}}{\varepsilon_{r2x}\varepsilon_{r2z}\sqrt{k_z^2 + k_x^2}} \quad (1)$$

with $\omega_2 \equiv c\sqrt{[k_x^2/(\varepsilon_{r2z}\mu_{r2y})] + [k_z^2/(\varepsilon_{r2x}\mu_{r2y})]}$, $\omega_1 = 2\pi f$, $k_x = -k cos(\theta_1)$, $k_z = k sin(\theta_1)$, $k = \omega_1/v_1$, $v_1 = c/\sqrt{\mu_{r1}\varepsilon_{r1}}$ and $c$ is the velocity of light in vacuum. From these expressions one may observe that, different from the case of the isotropic-to-isotropic temporal boundary, the incident angle $\theta_1$ of the monochromatic *p*-polarized wave is a key parameter to consider when changing $\varepsilon$ in time from isotropic to anisotropic. Now we ask: could we exploit such isotropic-to-anisotropic temporal boundary to achieve a *temporal Brewster angle*? would it be possible to eliminate the BW wave for a certain incident angle $\theta_I$ (for $t < t_1$) of the monochromatic *p*-polarized EM wave? To answer this question, in Eq. (1) we set the BW wave amplitude to zero and insert $\omega_2$ into Eq. (1), arriving at the following general expression:

$$\frac{\mu_{r2y}}{\varepsilon_{r2z}}\frac{\varepsilon_{r1}}{\mu_{r1}}\cos^2\theta_1 + \frac{\mu_{r2y}}{\varepsilon_{r2x}}\frac{\varepsilon_{r1}}{\mu_{r1}}\sin^2\theta_1 = 1 \qquad (2)$$

It is straightforward to notice that if $\varepsilon_{r2z} = \varepsilon_{r2x} = \varepsilon_{r2}$ (i.e., isotropic change) and $\mu_{r2y} \neq \mu_{r1}$ Eq. (2) is satisfied with $\frac{\varepsilon_{r1}}{\mu_{r1}} = \frac{\varepsilon_{r2}}{\mu_{r2y}}$, implying that the impedance of the medium before and after the temporal change of $\varepsilon$ and $\mu$ should be the same, as mentioned before, in order to achieve elimination of the BW wave. If we now consider the case of interest depicted in Fig. 1(d) with a constant $\mu_r$ and an isotropic-to-anisotropic change of $\varepsilon$ ($\varepsilon_{r2z} \neq \varepsilon_{r2x}$, $\mu_{r2y} = \mu_{r1}$) we obtain the following simple expression:

$$\theta_1 = \theta_{tB} = arcsin\left[\sqrt{\frac{(\varepsilon_{r2z}-\varepsilon_{r1})(\varepsilon_{r2x})}{(\varepsilon_{r2z}-\varepsilon_{r2x})(\varepsilon_{r1})}}\right] \qquad (3)$$

This simple, but important, closed-form expression describes our concept of *temporal Brewster angle* ($\theta_{tB}$). From Eq. (3), with $A \equiv \frac{(\varepsilon_{r2z}-\varepsilon_{r1})(\varepsilon_{r2x})}{(\varepsilon_{r2z}-\varepsilon_{r2x})(\varepsilon_{r1})}$, one may observe the following special cases where the *temporal Brewster angle* does not exist: I) for $\varepsilon_{r2x} < \varepsilon_{r2z} < \varepsilon_{r1}$ or $\varepsilon_{r1} < \varepsilon_{r2z} < \varepsilon_{r2x}$, $\sqrt{A}$ becomes imaginary, II) $\varepsilon_{r2z} = \varepsilon_{r2x}$ corresponding to the isotropic-to-isotropic temporal boundary, as depicted in Fig. 1(c) and finally if III) $\varepsilon_{r2z} = \varepsilon_{r1}$ or IV) $\varepsilon_{r2x} = \varepsilon_{r1}$, $A$ will become 0 or 1, respectively, meaning that $\theta_1 = 0°$ or $\theta_1 = 90°$, respectively. Cases I and II are obvious. As expected, no BW waves is produced for cases III and IV since there is no change of $\varepsilon$ for the component of the electric field that is present in the incident monochromatic wave ($E_z$ and $E_x$, respectively).

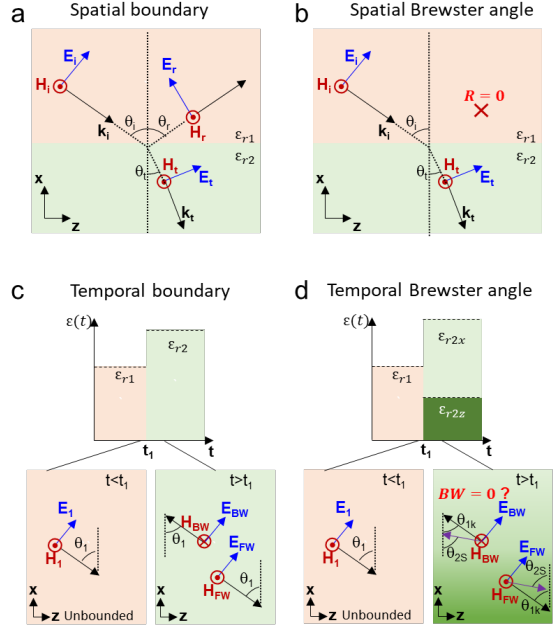

FIG. 1 (a) Spatial and (c) temporal boundaries. (b) Conventional (spatial) Brewster angle showing zero reflection for an obliquely incident *p*-polarized monochromatic wave. (d) *temporal Brewster angle*, considering a monochromatic *p*-polarized wave traveling in a time-dependent medium. The permittivity of the medium is changed in time from an isotropic $\varepsilon_{r1}$ to anisotropic $\overline{\overline{\varepsilon_{r2}}}=[\varepsilon_{r2x}, \varepsilon_{r2z}]$ at t = $t_1$. If $\theta_1$ is correctly selected to be *the temporal Brewster angle* the BW wave can be eliminated.

To further analyze the implications of Eq. (3), the analytical results of the amplitude of the FW and BW waves (calculated using Eq. (1)) considering an isotropic-to-anisotropic change of ε are shown in Fig. 2. Without loss of generality, here two cases are studied: $\varepsilon_{r1}$=10, $\varepsilon_{r2x} = 1$ with variable $\varepsilon_{r2z}$ (first row) and $\varepsilon_{r1}$=10, $\varepsilon_{r2z} = 1$ with variable $\varepsilon_{r2x}$ (second row). As observed in this figure, the amplitude of the BW wave is modified from a positive to a negative value depending on $\overline{\overline{\varepsilon_{r2}}}$ and incident angle $\theta_1$. The zero crossing (BW = 0) corresponds to the *temporal Brewster angle* (the analytical values for $\theta_{tB}$ calculated from Eq. (3) are shown as black dotted lines in Fig. 2(b,e) to guide the eye). For completeness, the theoretical results of the amplitude of the BW wave as a function of incidence angle for several cases of $\varepsilon_{r2z}$ and the amplitude of the BW wave as a function of $\varepsilon_{r2x}$ for several incident angles $\theta_1$ are shown in Fig. 2(c,f), respectively. (extracted from the vertical and horizontal white dashed lines in Fig. 2(b,e), respectively). From these results it is clear how a zero amplitude of the BW wave can be achieved by properly exploiting isotropic-to-anisotropic temporal modulation of ε and carefully selecting $\theta_1$ to be equal to what we call the *temporal the Brewster angle*. It is important to note that, unlike the spatial Brewster angle where the amplitude of the transmitted wave is maximized when using the spatial Brewster angle, the temporal version here shown will produce different amplitudes for the transmitted FW wave depending on the values of the permittivity tensor $\overline{\overline{\varepsilon_{r2}}}$, as expected from Eq. (1) and shown in Fig. 2 due to the induced temporal boundary at t = $t_1$. For instance, from Fig. 2(a-c) the FW wave has an amplitude of ~2 and ~1.67 when $\varepsilon_{r2z} = 15$ and $\varepsilon_{r2z} = 12.5$, respectively, considering that the oblique incidence angle coincides which their corresponding temporal Brewster angle ($\theta_{tB} = 10.8°$ and $\theta_{tB} = 8.4°$, respectively).

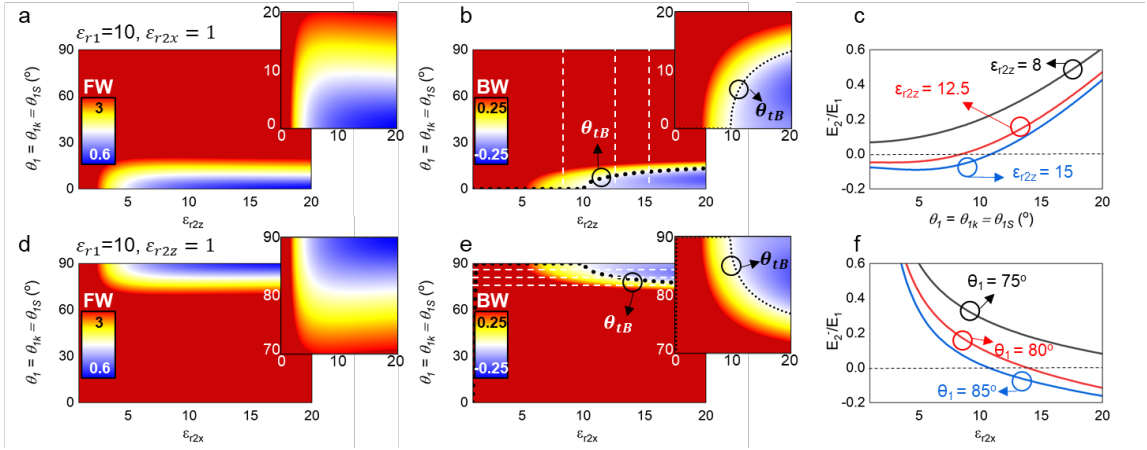

FIG. 2 Normalized amplitude of the forward wave (FW) (first column) and backward wave (BW) (second column) as a function of the incident angle $\theta_I$ and $\varepsilon_{r2z}$ (a,b) and $\varepsilon_{r2x}$ (d,e). The analytical values for the *temporal Brewster angle*, Eq. (3), are plotted in panels (b,e) as black dotted lines. (c,f) Normalized amplitude of the BW wave as a function of $\theta_I$ and $\varepsilon_{r2x}$ extracted from the white dotted lines in (b,e), respectively.

To demonstrate our *temporal Brewster angle* approach, we analytically calculate the magnetic field distribution ($H_y$) for the scenario shown as blue solid line in Fig. 2(c). We consider a monochromatic *p*-polarized wave traveling in a spatially unbounded medium with a time-dependent ε that is rapidly changed from isotropic $\varepsilon_{r1}$=10 to anisotropic $\overline{\overline{\varepsilon_{r2}}} = \{\varepsilon_{r2x} = 1, \varepsilon_{r2z} = 15\}$ at t = $t_1$. The analytical results for the incident (at a time t = $t_1^-$) and both FW and BW waves (at a time t = $t_1^+$) are shown in Fig. 3(a-c) considering three different incident angles: $\theta_I = 60°$, $\theta_I = 15°$ and $\theta_I = \theta_{tB} = 10.8°$, respectively. The Poynting vector (**S**) is also shown in each panel as black arrows demonstrating how **k** is preserved while the direction of the energy flow is modified when introducing an isotropic-to-anisotropic temporal boundary (in agreement with our previous findings [35]). Interestingly, note how the BW wave exist for incident angles $\theta_I = 60°$ and $\theta_I = 15°$ but it is eliminated when using the *temporal Brewster angle* as the incident angle ($\theta_I = \theta_{tB} = 10.8°$). Animations showing the configurations from Fig. 3(b,c) can be found as supplementary material.

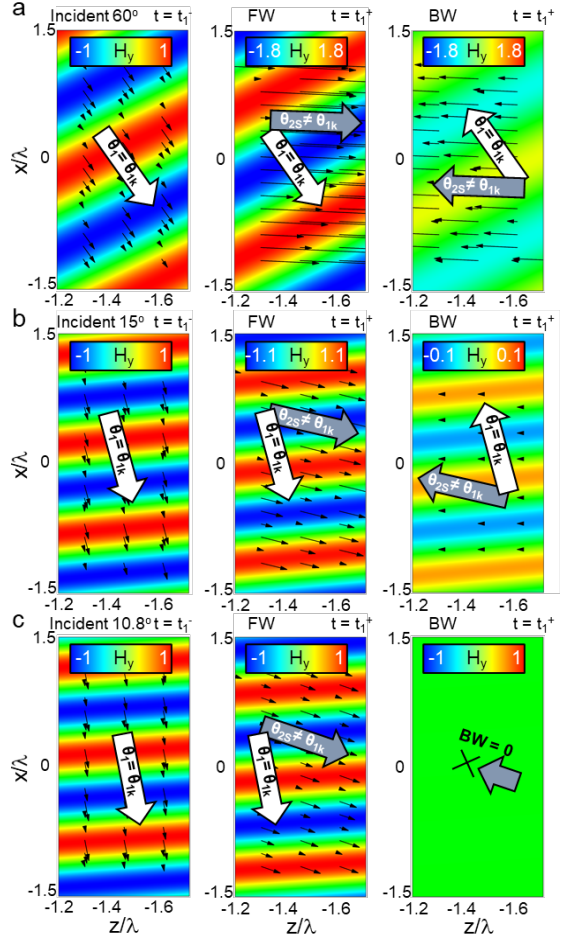

FIG. 3 Analytical results of the *y*-oriented magnetic field distribution for the incident (first column, t = $t_1^-$) and both FW and BW waves (second and third columns, t = $t_1^+$) considering a *p*-polarized monochromatic wave traveling in a medium whose $\varepsilon$ is rapidly changed in time from $\varepsilon_{r1} = 10$ to $\overline{\overline{\varepsilon_{r2}}} = \{\varepsilon_{r2x} = 1, \varepsilon_{r2z} = 15\}$. The incident angle is: (a) $\theta_I = 60°$, (b) $\theta_I = 15°$ and (c) $\theta_I = \theta_{tB} = 10.8°$.

To further corroborate the theoretical predictions, we numerically study the cases shown in Fig. 3(b,c) using the time-domain solver of the commercial software COMSOL Multiphysics® (following the setup from [35]) by considering an obliquely incident monochromatic *p*-polarized Gaussian beam. The numerical results for the out-of-plane magnetic field distribution before and after the change of ε from isotropic to anisotropic are shown in Fig. 4(a,b) for the same values of time-dependent ε as in Fig. 3(b,c), respectively, corroborating how the BW wave is eliminated when $\theta_I = \theta_{tB}$.

Finally, it is important to note that here we have focused our attention on the *p*-polarization. However, what will happen to an *s*-polarized wave? For the conventional spatial Brewster angle in Fig. 1(c) it is known how the *s*- polarization will be reflected at the interface between the two non-magnetic media (*x* = 0). In the temporal version different phenomena may occur: since an *s*-polarized monochromatic wave has only an out-of-plane component of the electric field ($E_y$) its performance will be similar to that of an isotropic-to-isotropic temporal change of ε (as shown in Fig. 1(c)) where incidence angle is irrelevant for the unbounded medium (with no spatial interface). Hence, an isotropic-to-anisotropic change of ε (in the *xz* plane) will only change the direction of propagation of the energy flow for a *p*-polarized wave. Additionally, if one only changes the in-plane components of the permittivity tensor $\overline{\overline{\varepsilon_{r2}}} = \{\varepsilon_{r2x}, \varepsilon_{r2z}\}$ no temporal boundary will be applied to the *s*-polarized wave, meaning that the *s*-polarized FW and BW waves will not be excited, i.e., the original incident wave will continue its propagation without being perturbed by the temporal change of ε. For the *s*-polarized BW and FW waves to exist one need to also change $\varepsilon_{r2y}$, as expected. For completeness, the numerical results of the out-of-plane electric field distribution of an oblique *s*-polarized Gaussian beam at different times are shown in Fig. 4 considering the same change of ε as in Fig. 4(b). As observed, the Gaussian beam does not change its direction of energy propagation (***S***) and no BW wave is produced for this polarization (see Supplementary Materials for an animation of this performance). Finally, it is important to note that here we have considered only changes of the permittivity tensor $\overline{\overline{\varepsilon_{r2}}} = \{\varepsilon_{r2x}, \varepsilon_{r2z}\}$. However, the direction of energy propagation ***S*** for the *s*-polarization may be changed in a similar fashion as the *p*-polarization when considering a change of the permeability tensor from an isotropic $\mu_1$ to an anisotropic permeability tensor $\overline{\overline{\mu_{r2}}} = \{\mu_{r2x}, \mu_{r2z}\}$ (not shown).

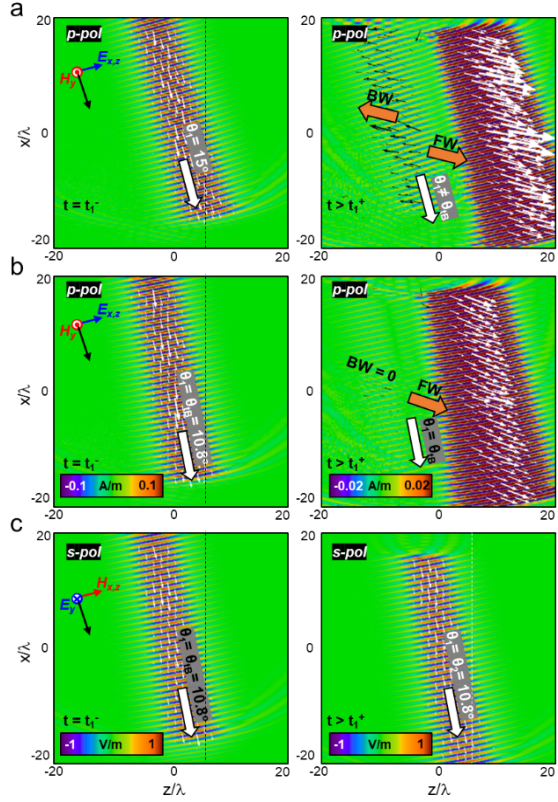

FIG. 4 Magnetic (a,b) and electric (c) field distributions for times t = $t_1^-$ (first column) and t > $t_1^+$ (second column) of a p- (a,b) and s-polarized (c) Gaussian beam traveling in a medium whose $\varepsilon$ is rapidly changed in time from $\varepsilon_{r1}$ = 10 to $\overline{\overline{\varepsilon_{r2}}}$ = {$\varepsilon_{r2x}$ = 1, $\varepsilon_{r2z}$ = 15}. The incidence angle is: (a) $\theta_1$ = 15° and (b,c) $\theta_1 = \theta_{tB}$ = 10.8°. Note that the source is switched off at t = $t_1$ and the color scale bar is chosen such that the $H_y$ field distributions for times t > $t_1^+$ have been saturated in order to clearly show and better appreciate the influence of the induced temporal boundaries on the BW wave.

In conclusion, we have shown, using analytical and numerical approaches, how the temporal analogue of the Brewster angle, which we have called *temporal Brewster angle,* can be achieved by considering a *p*-polarized monochromatic wave traveling in a temporal metamaterial whose permittivity is rapidly changed in time from isotropic to an anisotropic permittivity tensor. The physics behind the *temporal Brewster angle* have been presented and discussed and a closed-form theoretical expression has been derived. These findings have been analytically and numerically studied demonstrating how the BW wave is eliminated when the incidence angle equals the *temporal Brewster angle.* The results here presented can pave the way towards new and exciting ways of controlling polarization of waves and wave-matter interactions not only in space but also in time by exploiting temporal boundaries in temporal and spatiotemporal metamaterials.

V.P.-P. acknowledges support from the Newcastle University (Newcastle University Research Fellowship). N.E. would like to acknowledge the partial support from the Vannevar Bush Faculty

Fellowship program sponsored by the Basic Research Office of the Assistant Secretary of Defense for Research and Engineering, funded by the Office of Naval Research through grant N00014-16-1-2029.